\documentclass{revtex4}
\usepackage[ansinew]{inputenc}
\usepackage[dvips]{graphicx}
\usepackage{amsmath}
\usepackage{amsthm}
\usepackage{layout}
\usepackage{float}
\usepackage{subfigure}
\usepackage{amsfonts}
\usepackage{amssymb}

\begin{document}

\title{Measuring the Absolute Photo Detection Efficiency\\Using Photon Number Correlations}
\date{\today}
\begin{abstract}
We present two methods for determining the absolute detection efficiency of
photon-counting detectors directly from their singles rates under illumination
from a nonclassical light source. One method is based on a continuous variable
analogue to coincidence counting in discrete photon experiments, but does not
actually rely on high detector time resolutions. The second method is based on
difference detection which is a typical detection scheme in continuous
variable quantum optics experiments. Since no coincidence detection is
required with either method, they are useful for detection efficiency
measurements of photo detectors with detector time resolutions far too low to
resolve coincidence events.
\end{abstract}

\preprint{APS/123-QED}


\author{Michael Lindenthal and Johannes Kofler}%

\affiliation
{Institut f\"{u}r Experimentalphysik, Universit\"{a}%
t Wien, 1090 Wien, Austria \\Institut f\"{u}%
r Quantenoptik und Quanteninformation,\"{O}%
sterreichische Akademie der Wissenschaften, 1090 Wien, Austria}%

%
%
%
%
%
%
%
%
%
%
%
%
%
%
%
%
%
%
%
%
%
%
%
%
%
%
%
%
%
%


\maketitle

\section{Introduction}

High detection efficiency is crucial in many contexts within quantum optics.
In particular, recent linear optics quantum computation methods have been
shown to require detection efficiencies higher than $90$\% to be scalable~
\cite{knill_asfeqcwlo,obrien_doaaoqcng,gasparoni_roapcgsfqc,sanaka_enssfsqc}.
Another well known example comes from Bell inequality experiments where a
detection efficiency of at least $2/3$ is required to close the detection
loophole~\cite{eberhard_blacerfalfepre}. Thus higher detector efficiencies do
not only have the obvious advantage of providing more signal, but are also of
great relevance for both current quantum information science and fundamental
quantum physics. Therefore building new high efficiency detectors is a crucial
area of research. In this article we present methods to evaluate their
absolute detection efficiency. The typical way to measure detection efficiency
is to use a calibrated reference detector, and to compare its response to an
incident light beam of constant intensity with the response of the detector
under test. It has been shown that quantum mechanics itself provides a way to
measure the absolute detection efficiency of light detectors without the need
for a reference detector. By using non-classical photon statistics such as
those produced in spontaneous parametric down
conversion~\cite{kwiat_nhisopepp} both theory and experiment show that the
absolute detection efficiency can be determined by the ratio of the
coincidence rate to the singles
rate~\cite{klyshko_uotplfacopd,rarity_amodqeupd,kwiat_aeatrmospd}. However,
this method requires the detectors to have a high enough time resolution for
coincidence counting. We provide work-around schemes that overcome this
limitation. Specifically, we have developed two related detector efficiency
calibration methods that use quantum correlated light but do not rely on high
time resolution coincidence counting \cite{brida}. This might be of particular
interest for novel detectors and prototypes which are not yet capable of
coincidence counting, such as electron multiplying CCD cameras.

The first method is based on a continuous variable analogue of coincidence
counting in discrete photon experiments. Recall that coincidence counting is
essentially bitwise multiplication (i.e., an AND gate) of single counts within
a small coincidence window. In the continuous variable limit, this is achieved
by measuring the mean product of the detected photon numbers in two beams
generated by spontaneous parametric down conversion. Since down conversion
photons are emitted in pairs, the normalized mean product of the photon
numbers measured in the two beams during a specific time is maximal in the
case of perfect detection efficiencies when all photons are detected. On the
other hand, in the limit of small detection efficiencies, the normalized mean
product of the detected photon numbers has its minimum. For the general case
an expression can be derived, that allows calculation of detection
efficiencies from the measured mean product of the detected photon numbers in
the two down conversion beams.

The second method uses the variance of the detected photon number differences
in two beams generated by spontaneous parametric down conversion. Such
difference detection is a typical detection scheme in continuous variable
quantum optics experiments~\cite{heidmann_ooqnrotlb,aytuer_ptbol}. Since the
photons in the two beams are perfectly correlated, the difference in the
photon numbers measured in the two beams is zero in the case of perfect
detection efficiencies when all photons are detected. Uncorrelated loss in
these two beams diminishes those perfect correlations. Therefore in the limit
of small detection efficiencies, the normalized variance of the differences in
the detected photon numbers has its maximum. For the general case again an
expression can be derived, that allows to calculate the detection efficiencies
from the measured variance of the detected photon number differences in the
two down conversion beams.

The article is structured as follows: First we derive these relations
mentioned above disregarding background. Then we generalize these results to
include two different background levels in the two down conversion beams,
since background light is a very significant contribution in single photon
counting experiments.

\section{Theory}

\subsection{Product Detection Method}

\label{pd}

We begin by deriving a relationship between the mean product of the singles
rates and the detection efficiencies in each down conversion beam, $\eta_{1}$
and $\eta_{2}$, in the absence of background. We assume that parametric down
conversion emits light beams described by a general distribution $G_{N}(k)$ of
the number of photon pairs $k$ with mean value $\langle k\rangle_{G}=N$ and
the second moment $\langle k^{2}\rangle_{G}$~\cite{footnote}. The probability
of detecting $l$ out of $k$ photons in each of the two down conversion beams
is given by the binomial distribution $B_{k,\eta_{i}}(l)={\binom{k}{l}}%
\,\eta_{i}^{l}\,(1-\eta_{i})^{k-l}$, with mean value $\langle l\rangle
_{B}=\eta_{i}k$ and the second moment $\langle l^{2}\rangle_{B}=\eta_{i}%
k-\eta_{i}^{2}k+\eta_{i}^{2}k^{2}$ ($i=1,2$). Thus the mean product of the
detected photon numbers in the two down conversion beams is given by
\begin{align}
\langle lm\rangle &  =\sum_{k=0}^{\infty}\sum_{l=0}^{k}\sum_{m=0}^{k}%
G_{N}(k)\,B_{k,\eta_{1}}(l)\,B_{k,\eta_{2}}(m)\,l\,m\nonumber\\
&  =\eta_{1}\eta_{2}\langle k^{2}\rangle_{G}=\frac{\eta_{2}}{\eta_{1}}(\langle
l^{2}\rangle-\langle l\rangle+\eta_{1}\langle l\rangle)\,,
\end{align}
where $l$ and $m$ are the detected photon numbers in each down conversion
beam, $\langle l\rangle=\eta_{1}\langle k\rangle_{G}$ and $\langle
l^{2}\rangle=\eta_{1}\langle k\rangle_{G}-\eta_{1}^{2}\langle k\rangle
_{G}+\eta_{1}^{2}\langle k^{2}\rangle_{G}$.

Together with the expression $\frac{\langle l\rangle}{\langle m\rangle}%
=\frac{\eta_{1}}{\eta_{2}}$, the detection efficiency $\eta_{1}$ follows as
\begin{equation}
\eta_{1}=\frac{\langle lm\rangle}{\langle m\rangle}-\frac{\langle l^{2}%
\rangle}{\langle l\rangle}+1\,. \label{res1}%
\end{equation}
The corresponding result for $\eta_{2}$ is
\begin{equation}
\eta_{2}=\frac{\langle lm\rangle}{\langle l\rangle}-\frac{\langle m^{2}%
\rangle}{\langle m\rangle}+1\,. \label{res2}%
\end{equation}
Note that neither formula depends on the coincidence rate. However, the
quantum statistics of the light enters the expressions in the mean product of
the singles rates.

In a former method~\cite{klyshko_uotplfacopd} the absolute detection
efficiency is determined from the ratio of the mean coincidence rate $\langle
c\rangle$ to the mean singles rate $\langle l\rangle$ or $\langle m\rangle$,
$\eta_{1}=\frac{\langle c\rangle}{\langle m\rangle}$ and $\eta_{2}%
=\frac{\langle c\rangle}{\langle l\rangle}$. Together with (\ref{res1}) or
(\ref{res2}) one obtains for $\eta=\eta_{1}=\eta_{2}$, and hence $\langle
s\rangle=\langle l\rangle=\langle m\rangle$ and $\langle s^{2}\rangle=\langle
l^{2}\rangle=\langle m^{2}\rangle$,
\begin{equation}
\langle s\rangle-\langle c\rangle=\langle s^{2}\rangle-\langle lm\rangle
=\frac{\langle(l-m)^{2}\rangle}{2}\,. \label{eq 4}%
\end{equation}
This simple expression relates the difference between the mean singles rate
and the mean coincidence rate to the variance of the detected photon number
differences in the two down conversion beams. This motivates our second
approach for determining the detection efficiencies from the variance of the
detected photon number differences in the two down conversion beams described
in section \ref{dd}.

\subsection{Product Detection Method: General Approach Including Background}

\label{pde}

We now extend this theory to cover more realistic experimental conditions and
correct the measurements for possibly different backgrounds in the two
detectors. The averaged quantities contained in (\ref{res1}) and (\ref{res2})
have to be extracted from experimentally accessible quantities which include
background. We do so by splitting up the measured photon numbers (subscript M)
into the photon numbers corresponding to the signal (no subscript) and into
photon numbers corresponding to the background (subscript B), where
$l=l_{M}-l_{B}$ and $m=m_{M}-m_{B}$. The background can be estimated
experimentally from a separate configuration. We get%
\begin{align}
\langle l\rangle &  =\langle l_{M}\rangle-\langle l_{B}\rangle\,,\label{bg1}\\
\langle m\rangle &  =\langle m_{M}\rangle-\langle m_{B}\rangle\,,\\
\langle l^{2}\rangle &  =\langle l_{M}^{2}\rangle-\langle l_{B}^{2}%
\rangle-2\,\langle l_{M}\rangle\langle l_{B}\rangle+2\,\langle l_{B}%
\rangle^{2}\,,\\
\langle m^{2}\rangle &  =\langle m_{M}^{2}\rangle-\langle m_{B}^{2}%
\rangle-2\,\langle m_{M}\rangle\langle m_{B}\rangle+2\,\langle m_{B}%
\rangle^{2}\,, \label{bg2}%
\end{align}
and
\begin{equation}
\langle lm\rangle=\langle l_{M}m_{M}\rangle-\langle l_{M}\rangle\langle
m_{B}\rangle-\langle l_{B}\rangle\langle m_{M}\rangle+\langle l_{B}%
\rangle\langle m_{B}\rangle\,.
\end{equation}
Here we used the statistical independence of $l$ and $l_{B}$ and $m$ and
$m_{B}$, respectively. By inserting these expressions into (\ref{res1}) and
(\ref{res2}) the detection efficiencies can be determined from the data
directly measurable in an experiment:
\begin{equation}
\eta_{1}=\frac{\langle l_{M}m_{M}\rangle-\langle l_{M}\rangle\langle
m_{B}\rangle-\langle l_{B}\rangle\langle m_{M}\rangle+\langle l_{B}%
\rangle\langle m_{B}\rangle}{\langle m_{M}\rangle-\langle m_{B}\rangle}%
-\frac{\langle l_{M}^{2}\rangle-\langle l_{B}^{2}\rangle-2\,\langle
l_{M}\rangle\langle l_{B}\rangle+2\,\langle l_{B}\rangle^{2}}{\langle
l_{M}\rangle-\langle l_{B}\rangle}+1
\end{equation}
and
\begin{equation}
\eta_{2}=\frac{\langle l_{M}m_{M}\rangle-\langle l_{M}\rangle\langle
m_{B}\rangle-\langle l_{B}\rangle\langle m_{M}\rangle+\langle l_{B}%
\rangle\langle m_{B}\rangle}{\langle l_{M}\rangle-\langle l_{B}\rangle}%
-\frac{\langle m_{M}^{2}\rangle-\langle m_{B}^{2}\rangle-2\,\langle
m_{M}\rangle\langle m_{B}\rangle+2\,\langle m_{B}\rangle^{2}}{\langle
m_{M}\rangle-\langle m_{B}\rangle}+1\,.
\end{equation}

\subsection{Difference Detection Method}

\label{dd}

Our second approach for determining the absolute detection efficiencies in
each down conversion beam, $\eta_{1}$ and $\eta_{2}$, relies on measuring the
variance of the differences in the singles rates $\langle(l-m)^{2}\rangle$.
This method is closely related to the approach described in section \ref{pd},
since $\langle(l-m)^{2}\rangle=\langle l^{2}\rangle+\langle m^{2}%
\rangle-2\,\langle lm\rangle$, and both $\langle(l-m)^{2}\rangle$ and $\langle
lm\rangle$ depend on the degree of second-order coherence which is affected by
uncorrelated loss. However, the two methods may be useful under different
circumstances, especially since difference detection is a typical detection
scheme in continuous variable quantum optics experiments.

The variance of the detected photon number differences in the two down
conversion beams is given by
\begin{align}
\langle(l-m)^{2}\rangle &  =\sum_{k=0}^{\infty}\sum_{l=0}^{k}\sum_{m=0}%
^{k}G_{N}(k)\,B_{k,\eta_{1}}(l)\,B_{k,\eta_{2}}(m)\,(l-m)^{2}\nonumber\\
&  =\langle l\rangle+\langle m\rangle-(\eta_{1}\langle l\rangle+\eta
_{2}\langle m\rangle)+(\eta_{1}-\eta_{2})^{2}\frac{\langle l^{2}%
\rangle-\langle l\rangle+\eta_{1}\langle l\rangle}{\eta_{1}^{2}}\,,
\label{noisegen}%
\end{align}
where $l$ and $m$ are the detected photon numbers in each down conversion beam.

Together with the expression $\frac{\langle l\rangle}{\langle m\rangle}%
=\frac{\eta_{1}}{\eta_{2}}$, the detection efficiency $\eta_{1}$ follows as
\begin{equation}
\eta_{1}=\frac{3\,\langle m\rangle-\frac{\langle m\rangle^{2}}{\langle
l\rangle}+\langle l^{2}\rangle(1-\frac{\langle m\rangle}{\langle l\rangle
})^{2}-\langle(l-m)^{2}\rangle}{2\,\langle m\rangle}\,. \label{res3}%
\end{equation}
Correspondingly,
\begin{equation}
\eta_{2}=\frac{3\,\langle l\rangle-\frac{\langle l\rangle^{2}}{\langle
m\rangle}+\langle m^{2}\rangle(1-\frac{\langle l\rangle}{\langle m\rangle
})^{2}-\langle(l-m)^{2}\rangle}{2\,\langle l\rangle}\,. \label{res4}%
\end{equation}
\newline

For $\eta=\eta_{1}=\eta_{2}$, and hence $\langle s\rangle=\langle
l\rangle=\langle m\rangle$, (\ref{noisegen}) becomes%
\begin{equation}
\langle(l-m)^{2}\rangle=2\,(1-\eta)\langle s\rangle\,. \label{noises}%
\end{equation}
For perfect detection efficiencies ($\eta=1$), $\langle(l-m)^{2}\rangle=0$.
Since the down-converted photons are always created in pairs, the difference
in the detected photon number, and hence the variance of that quantity, is
exactly zero if all of the pairs are detected. In the limit of very small
detection efficiencies ($\eta\ll1$), $\langle(l-m)^{2}\rangle\approx2\langle
s\rangle$, which corresponds to the variance of two independent Poissonian
light beams of equal intensities. The non-Poissonian contributions in
$G_{N}(k)$ cancel out, showing that quantum statistics strongly depends on the
detection efficiency of the detectors.

Expression (\ref{noises}) can be rewritten as%
\begin{equation}
\eta=1-\frac{\langle(l-m)^{2}\rangle}{2\,\langle s\rangle}\,. \label{eq tja}%
\end{equation}
A similar expression was stated without explicit derivation
previously~\cite{heidmann_ooqnrotlb,aytuer_ptbol}.

With $\langle s\rangle=\eta\,N$, the normalized expression (\ref{noises})
reads $\frac{\langle(l-m)^{2}\rangle}{\langle s\rangle^{2}}=\frac{2}{N}%
(\frac{1}{\eta}-1)$, which verifies that the normalized variance of the
detected photon number differences in the two down conversion beams diverges
for $\eta\rightarrow0$ and goes to zero for $\eta\rightarrow1$. Using equation
(\ref{eq 4}), an analogous treatment can be performed for the normalized mean
product of the detected photon numbers in the two beams.

\subsection{Difference Detection Method: General Approach Including Background}

As in section \ref{pde}, we extend this theory to cover the more realistic
experimental conditions including background. The quantities in (\ref{res3})
and (\ref{res4}) have to be extracted from quantities that are directly
accessible to measurement. We get (\ref{bg1})--(\ref{bg2}) and%
\begin{equation}
\langle(l-m)^{2}\rangle=\langle(l_{M}-m_{M})^{2}\rangle+2\,(\langle
l_{B}\rangle-\langle m_{B}\rangle)^{2}-2\,(\langle l_{M}\rangle-\langle
m_{M}\rangle)(\langle l_{B}\rangle-\langle m_{B}\rangle)+2\,\langle
l_{B}\rangle\langle m_{B}\rangle-\langle l_{B}^{2}\rangle-\langle m_{B}%
^{2}\rangle\label{exp}%
\end{equation}
for the background corrected difference term.

By inserting (\ref{bg1})--(\ref{bg2}) and (\ref{exp}) into (\ref{res3}) and
(\ref{res4}) the detection efficiencies can be determined from the data
directly measurable in an experiment,%
\begin{align}
\eta_{1}  &  =\frac{3\,(\langle m_{M}\rangle-\langle m_{B}\rangle
)-\frac{(\langle m_{M}\rangle-\langle m_{B}\rangle)^{2}}{\langle l_{M}%
\rangle-\langle l_{B}\rangle}}{2\,(\langle m_{M}\rangle-\langle m_{B}\rangle
)}+\frac{\langle l_{M}^{2}\rangle-\langle l_{B}^{2}\rangle-2\,\langle
l_{M}\rangle\langle l_{B}\rangle+2\langle l_{B}\rangle^{2}}{2\,(\langle
m_{M}\rangle-\langle m_{B}\rangle)}\left(  1-\frac{\langle m_{M}%
\rangle-\langle m_{B}\rangle}{\langle l_{M}\rangle-\langle l_{B}\rangle
}\right)  ^{2}\nonumber\\
&  \quad-\frac{\langle(l_{M}-m_{M})^{2}\rangle+2\,(\langle l_{B}%
\rangle-\langle m_{B}\rangle)^{2}}{2\,(\langle m_{M}\rangle-\langle
m_{B}\rangle)}-\frac{2\,(\langle l_{M}\rangle-\langle m_{M}\rangle)(\langle
l_{B}\rangle-\langle m_{B}\rangle)}{2\,(\langle m_{M}\rangle-\langle
m_{B}\rangle)}+\frac{2\,\langle l_{B}\rangle\langle m_{B}\rangle-\langle
l_{B}^{2}\rangle-\langle m_{B}^{2}\rangle}{2\,(\langle m_{M}\rangle-\langle
m_{B}\rangle)}\,.
\end{align}
and
\begin{align}
\eta_{2}  &  =\frac{3\,(\langle l_{M}\rangle-\langle l_{B}\rangle
)-\frac{(\langle l_{M}\rangle-\langle l_{B}\rangle)^{2}}{\langle m_{M}%
\rangle-\langle m_{B}\rangle}}{2\,(\langle l_{M}\rangle-\langle l_{B}\rangle
)}+\frac{\langle m_{M}^{2}\rangle-\langle m_{B}^{2}\rangle-2\,\langle
m_{M}\rangle\langle m_{B}\rangle+2\langle m_{B}\rangle^{2}}{2\,(\langle
l_{M}\rangle-\langle l_{B}\rangle)}\left(  1-\frac{\langle l_{M}%
\rangle-\langle l_{B}\rangle}{\langle m_{M}\rangle-\langle m_{B}\rangle
}\right)  ^{2}\nonumber\\
&  \quad-\frac{\langle(l_{M}-m_{M})^{2}\rangle+2(\langle l_{B}\rangle-\langle
m_{B}\rangle)^{2}}{2\,(\langle l_{M}\rangle-\langle l_{B}\rangle)}%
-\frac{2\,(\langle l_{M}\rangle-\langle m_{M}\rangle)(\langle l_{B}%
\rangle-\langle m_{B}\rangle)}{2\,(\langle l_{M}\rangle-\langle l_{B}\rangle
)}+\frac{2\,\langle l_{B}\rangle\langle m_{B}\rangle-\langle l_{B}^{2}%
\rangle-\langle m_{B}^{2}\rangle}{2\,(\langle l_{M}\rangle-\langle
l_{B}\rangle)}\,.
\end{align}

\subsection{Error Estimates\label{sect err}}

Finally, we want to derive the statistical errors of the detection
efficiencies (without background), i.e., ($A$) for the product detection,
eq.~(\ref{res1}), ($B$) for the difference detection, eq.~(\ref{res3}), and
($C$) for the coincidence method~\cite{klyshko_uotplfacopd}, described above
eq.~(\ref{eq 4}):%
\begin{align}
\eta_{1}^{(A)}  &  =\frac{\langle lm\rangle}{\langle m\rangle}-\frac{\langle
l^{2}\rangle}{\langle l\rangle}+1\,,\\
\eta_{1}^{(B)}  &  =\frac{3\,\langle m\rangle-\frac{\langle m\rangle^{2}%
}{\langle l\rangle}+\langle l^{2}\rangle(1-\frac{\langle m\rangle}{\langle
l\rangle})^{2}-\langle(l-m)^{2}\rangle}{2\,\langle m\rangle}\,,\\
\eta_{1}^{(C)}  &  =\frac{\langle c\rangle}{\langle m\rangle}\,,
\end{align}
In each case $\eta_{1}$ is a function of several mean values, i.e., $\eta
_{1}=\eta_{1}(\langle u\rangle,\langle v\rangle,...)$ where $u,v,...\in
\{l,m,l^{2},lm,(l-m)^{2},c\}$. The sample variance of $\eta_{1}$ is defined as%
\begin{equation}
\sigma^{2}(\eta_{1})\equiv\sigma_{\langle u\rangle}^{2}\left(  \frac
{\partial\eta_{1}}{\partial\langle u\rangle}\right)  ^{\!2}+\sigma_{\langle
v\rangle}^{2}\left(  \frac{\partial\eta_{1}}{\partial\langle v\rangle}\right)
^{\!2}+2\,\sigma_{\langle u\rangle\langle v\rangle}\,\frac{\partial\eta_{1}%
}{\partial\langle u\rangle}\,\frac{\partial\eta_{1}}{\partial\langle v\rangle
}+...
\end{equation}
where%
\begin{align}
\sigma_{\langle u\rangle}^{2}  &  \equiv\frac{\langle u^{2}\rangle-\langle
u\rangle^{2}}{M}\,,\\
\sigma_{\langle u\rangle\langle v\rangle}  &  \equiv\frac{\langle
uv\rangle-\langle u\rangle\langle v\rangle}{M}%
\end{align}
are the variances and covariances of the sample means with sample size $M$. It
has to be stressed that in experiments the time interval chosen for
accumulating the individual measurements needs to be much larger than the
resolving time of the detector under test. The mean values are given by%
\begin{equation}
\langle x\rangle=\sum_{k=0}^{\infty}\sum_{l=0}^{k}\sum_{m=0}^{k}%
G_{N}(k)\,B_{k,\eta_{1}}(l)\,B_{k,\eta_{2}}(m)\,x\,.
\end{equation}
Due to the perfect correlations the first and second moment of the
coincidences can be computed by applying the binomial distribution to one down
conversion arm twice:%
\begin{equation}
\langle c^{p}\rangle=\sum_{k=0}^{\infty}\sum_{l=0}^{k}\sum_{m=0}^{l}%
G_{N}(k)\,B_{k,\eta_{1}}(l)\,B_{k,\eta_{2}}(m)\,m^{p}\,,
\end{equation}
where $p=1,2$.

For a Poissonian down conversion distribution $G_{N}(k)=N^{k}\exp(-N)/k!$,
with $N$ the expected mean photon number in one sample measurement, the
efficiency sample variances in the three cases are%
\begin{align}
\sigma^{2}(\eta_{1}^{(A)})  &  =\frac{\eta_{1}\,(1+N-\eta_{1})+N\,\eta
_{2}\,[2+\eta_{1}\,(\eta_{1}-4)]}{M\,N\,\eta_{2}}\,,\\
\sigma^{2}(\eta_{1}^{(B)})  &  =\frac{2\,\eta_{1}^{4}\,(N\,\eta_{2}%
-1)+2\,\eta_{1}^{3}\,[1+N\,(1+2\,\eta_{2}\,(\eta_{2}-3))]+N\,\eta_{1}%
^{2}\,\eta_{2}\,[5-2\,\eta_{2}(\eta_{2}-2)]-4\,N\,\eta_{1}\,\eta_{2}%
^{2}+N\,\eta_{2}^{3}}{2\,M\,N\,\eta_{1}^{2}\,\eta_{2}}\,,\\
\sigma^{2}(\eta_{1}^{(C)})  &  =\frac{\eta_{1}\,(1+\eta_{1}-2\,\eta_{1}%
\,\eta_{2})}{M\,N\,\eta_{2}}\,.
\end{align}
In the limit $N\gg1$ the variance for the coincidence method
vanishes:\ $\sigma^{2}(\eta_{1}^{(C)})\rightarrow0$. For the product and
difference detection the variances approach constant values which depend on
the efficiencies $\eta_{1}$ and $\eta_{2}$ as well as the sample size $M$.
Figure 1 shows these two variances as a function of $\eta_{1}$ for $M=1$ where
all sample variances scale inversely with the sample size. In general, the
difference method is more accurate than the product method, except in the case
of fixed $\eta_{2}$ and vanishing $\eta_{1}$. In the special case of equal
efficiencies $\eta_{1}=\eta_{2}$ the expression for the difference method
simplifies tremendously:%
\begin{equation}
\sigma^{2}(\eta_{1}^{(B)})|_{\eta_{2}=\eta_{1}}=\frac{2\,(1-\eta_{1})^{2}}%
{M}\,. \label{var Poiss}%
\end{equation}
\begin{figure}[t]
\begin{center}
\includegraphics{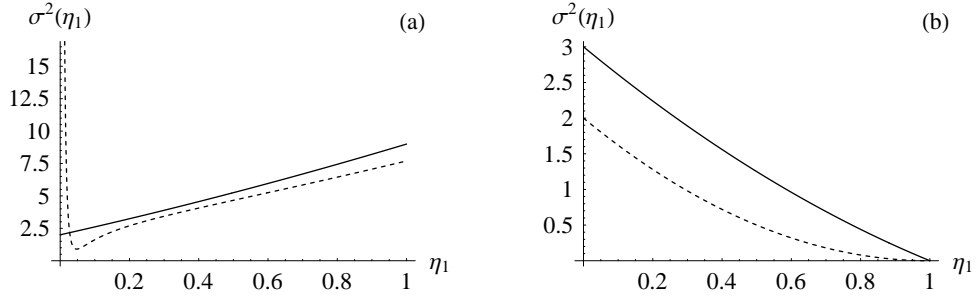}
\end{center}
\par
\vspace{-0.25cm} \caption{Sample variances $\sigma^{2}(\eta_{1}^{(A)})$
(product method, solid line) and $\sigma^{2}(\eta_{1}^{(B)})$ (difference
method, dashed line) as a function of $\eta_{1}$ for a Poissonian distribution
in the limit $N\gg1$. For the sake of generality the sample size is $M=1$,
where all\ variances scale with $1/M$. (a) The second efficiency is
constant:\ $\eta_{2}=0.1$. The variance $\sigma^{2}(\eta_{1}^{(B)})$ diverges
for $\eta_{1}\rightarrow0$. (b) Equal efficiencies:\ $\eta_{2}=\eta_{1}$.}%
\label{fig variance}%
\end{figure}

For a thermal distribution $G_{N}(k)\propto\exp(-k/N)$ the first two variances
are more cumbersome and we do not write them here. In the limit of increasing
$N$ the variance for the coincidence method $\sigma^{2}(\eta_{1}^{(C)})$
vanishes again, while the variances for the product and difference detection
method linearly diverge in the limit $N\gg1$, still also scaling with $1/M$.
Only in the special case of equal efficiencies $\eta_{1}=\eta_{2}$ the
variance for the difference method $\sigma^{2}(\eta_{1}^{(B)})$ becomes
independent of $N$. Therefore, if equal detectors are used, the difference
method is more favorable. In this special case and for $N\gg1$ we have the
simple expression%
\begin{equation}
\sigma^{2}(\eta_{1}^{(B)})|_{\eta_{2}=\eta_{1}}=\frac{4\,(1-\eta_{1})^{2}}%
{M}\,.
\end{equation}
Hence, in this case, the variance for the thermal distribution has the same
form as expression (\ref{var Poiss}) for the Poissonian distribution.

\section{Conclusions and Outlook}

We have presented two methods for determining the absolute detection
efficiency of photo detectors. Since they are applicable to detectors with low
time resolution, they overcome the limitations typical for absolute detection
efficiency measurements. The first is based on measuring the mean product of
the detected singles rates in two beams generated by spontaneous parametric
down conversion. The second method uses the variance measurements of the
differences in the detected singles rates in the two down conversion beams.
The two methods correspond to the different detection methods typically used
in either the discrete photon or continuous variable communities,
respectively. Both procedures could be used for measuring the absolute
detection efficiency of photo detectors that do not provide the appropriate
time resolution for coincidence counting.

\begin{acknowledgments}
This work was funded by the DOC-program of the Austrian Academy of
Sciences as well as the Austrian Science Foundation (FWF), Project
SFB 1506. We thank Kevin Resch, Markus Aspelmeyer, Aephraim
Steinberg, Andrew White and Gregor Weihs for helpful and
motivating discussions. Especially we thank Anton Zeilinger for
supporting this work.
\end{acknowledgments}

\appendix

\end{document}